\documentclass[%
 reprint,
 amsmath,amssymb,
 aps,
]{revtex4-2}

\usepackage{braket}
\usepackage{graphicx}% Include figure files
\usepackage{dcolumn}% Align table columns on decimal point
\usepackage{bm}% bold math
\usepackage{chemformula}
\begin{document}

\preprint{APS/123-QED}

\title{Degree of atomicity in the chemical bonding: Why to return to the \ch{H2} molecule?}% Force line breaks with \\

\author{Maciej Hendzel}
\email{maciej.hendzel@doctoral.uj.edu.pl}
\author{J\'ozef Spa\l{}ek}
\email{jozef.spalek@uj.edu.pl (corresponding author)}
 \affiliation{Institute of Theoretical Physics, Jagiellonian University,\\ ul.~\L{}ojasiewicza 11, PL-30-348 Krak\'{o}w, Poland}%Lines break automatically or can be forced with \\

\date{\today}% It is always \today, today,
             %  but any date may be explicitly specified

\begin{abstract}

We analyze two-particle binding factors for the case of \ch{H2} molecule with the help of our original Exact Diagonalization \textit{Ab Intio} (EDABI) approach. Explicitly, we redefine the
many-particle covalency and ionicity factors as a function of interatomic distance. Insufficiency of those basic characteristics is stressed and the
concept of \textit{atomicity} is introduced and corresponds to the Mott and Hubbard criteria
concerning the localization in many-particle systems. This additional characteristic
introduces atomic ingredient into the essentially molecular states and thus eliminates a spurious
behavior of the standard covalency factor with the increasing interatomic distance, as well as provides
a physical reinterpretation of the chemical bond's nature.
\end{abstract}

%\keywords{Suggested keywords}%Use showkeys class option if keyword
                              %display desired
\maketitle

%\tableofcontents

\section{Introduction and motivation}

The concept of chemical bond as the fundamental quantum--mechanical characteristic of molecules 
such as \ch{H2}, was firmly established by Heitler--London \cite{HeitlerLondon} in 1927. This 
pioneering quantitative paper was based, by today's standards, on the Hartree--Fock 
approximation for the two--particle wave function of the two electrons in \ch{H2} molecule. Later, this function has been expressed by 
the corresponding atomic $1s$ hydrogen wave functions in the form
of symmetrized product with antisymmetrized spin part, the latter reflecting the spin--singlet 
ground state. Such a selection of the component atomic wave functions, represented a rather drastic approximation and has been corrected subsequently by 
selecting their superposition of those atomic wave functions into molecular 
single--particle wave functions centered on individual atoms, which have been subsequently put 
into a proper two--particle form \cite{Slater}. This whole procedure established a 
canonical viewpoint of the \textit{covalent} bond, with a degree of ionicity (double occupancy of 
individual atoms) introduced \textit{ad hoc} later to it (Valence Bond Theory) \cite{Slater}. Theory of the bonding reached its mature form with an excellent series of papers by Ko\l{}os nad Wolniewicz \cite{KW1963,KW1968}
who have included higher (virtually) excited states, supplemented with the nuclear vibrations 
\cite{KolosVib} to a fully quantitative form, which has been subsequently tested experimentally, since the 
bonding in \ch{H2} molecule represents one of the tests of 
quantum--mechanical--theory verification in quantum chemistry \cite{Herzberg}.

In this brief paper we address, first of all, the question why we must realize that there is a need to return to the problem origins of the bonding nature in the \ch{H2} molecule. Namely, we have observed recently that the two--electron wave function, representing the single bond, composed of originally $1s$ electrons of hydrogen atoms contains an inherent inconsistency when we interpret covalency in the standard manner \cite{Hendzel1, Hendzel2}. Explicitly, when starting from \textit{an exact solution} of the Heitler--London problem (with proper molecular single--particle wave functions included at the start), we have detected that the covalency increases with the increasing distance between the nuclei, a clearly unphysical feature. As a subsidiary observation we have noted that the Heitler--London (Hartree--Fock) two--electron wave function leads to nonzero (actually, maximal) value of covalency in the limit of entirely separated atoms. Such an inconsistency has brought to our attention the old concept of Mott \cite{Mott}, concerning the electron localization in condensed matter physics (see also \cite{PRLSpal, SpalJul}). In effect, we have decided to introduce the concept of \textit{atomicity} in the context of the correlated molecular electronic states \cite{Hendzel1}. This concept represents a novel nontrivial feature of the chemical bond, since it is introduced as an external factor into an essentially molecular (collective) language of the covalent bonding, including also the ionicity. Hence, in this paper we summarize and mainly interpret our recent results \cite{Hendzel1, Hendzel2} which, in our view, provide a connection between (correlated) states of small molecules and condensed matter physics, as well as delineate the essential difference between the two. 

The structure of this paper is as follows. In the next Section we briefly summarize our method and in Sec. III, regarded as the main part, we discuss our results and their meaning. This is followed by a brief Outlook. In general, the aim of the paper is to supplement previous papers \cite{Hendzel2, Hendzel1} with detailed discussion and interpretation of the results. Such a discussion may be of importance when the concept of \textit{atomicity} is analyzed for more complicated bonds such as C--C in the hydrocarbons. The connecting link between the condensed matter localization and \textit{molecular atomicity} may be then applied also to other nano--systems \cite{SpalJul}.

\section{Method}

Our approach is based on \textbf{E}xact \textbf{D}iagonalization\textbf{ Ab I}nitio (EDABI) method which has been proposed and developed in our group \cite{JS2000, Rycerz2001}. Here we use this method to provide complementary bonding characteristics on example of \ch{H2} molecule. The starting Hamiltonian, containing all Coulomb interactions, formulated in the second quantization language, is of the form
\begin{align}
     &\mathcal{\hat{H}} = \epsilon_a \sum_i \hat{n}_{i\sigma} 
    + {\sum_{ij\sigma}}' t_{ij}\,\hat{a}^{\dag}_{i\sigma} \,\hat{a}_{j\sigma} + U \sum_{i} \hat{n}_{i\uparrow}\,\hat{n}_{i\downarrow} \nonumber \\ &+ \frac{1}{2} {\sum_{ij}}' K_{ij}\hat{n}_{i}\,\hat{n}_{j}
     -  \frac{1}{2}{\sum_{ij}}' J^H_{ij}  \left(\hat{\textbf{S}}_i \cdot \hat{\textbf{S}}_j-\frac{1}{4}
    \hat{n}_i\hat{n}_j\right) \nonumber \\ &+ \frac{1}{2} {\sum_{ij}}' {J}'_{ij}
    (\hat{a}^{\dagger}_{i\uparrow}\hat{a}^{\dagger}_{i\downarrow}\hat{a}_{j\downarrow}\hat{a}_{j\uparrow} + \mathrm{H.c.})  \nonumber \\
     &+\frac{1}{2} {\sum_{ij}}' V_{ij} (\hat{n}_{i\sigma}+\hat{n}_{j\sigma})(\hat{a}^{\dagger}_{i\bar{\sigma}}\hat{a}_{j\bar{\sigma}}+ \mathrm{H.c.}) + \mathcal{H}_{\text{ion-ion}},
    \label{Hamiltonian_eq}
\end{align}

\noindent
where H.c. denotes the Hermitian conjugation, $\hat{a}_{i\sigma}$ ($\hat{a}^\dagger_{i\sigma}$) are fermionic annihilation (creation) operators for state $i$ and spin $\sigma$, $\hat{n}_{i\sigma} \equiv \hat{a}^\dagger_{i\sigma} \hat{a}_{i\sigma}$, and $\hat{n}_i \equiv \hat{n}_{i\uparrow} + \hat{n}_{i\downarrow} \equiv
\hat{n}_{i\sigma}+ \hat{n}_{i\bar{\sigma}}$. The spin operators are defined as $\hat{\textbf{S}}_i \equiv \frac{1}{2} \sum_{\alpha\beta} \hat{a}^\dagger_{i\alpha} \bm{\sigma}_i^{\alpha\beta} \hat{a}_{i\beta}$ with $\sigma_i$ representing Pauli matrices. The primed summations mean that $i\neq j$. The Hamiltonian contains the atomic and hopping parts ($\propto \epsilon_a$ and $t_{ij}$, respectively), the so-called Hubbard term $\propto U$; representing the intra-atomic interaction between the particles on the same atomic site \emph{i} with opposite spins, the direct intersite Coulomb interaction $\propto K_{ij}$, Heisenberg exchange $\propto J^H_{ij}$, and the two-particle and the correlated hopping terms ($\propto J_{ij}^\prime$ and $V_{ij}$, respectively). The last term describes the ion-ion Coulomb interaction which is adopted here in its classical form.

By way of diagonalization of Hamiltonian (Eq. \eqref{Hamiltonian_eq}) one can write ground state energy with the ground--state two--particle wave function, obtained in the form $\psi_G(\textbf{r}_1, \textbf{r}_2) = \psi_{cov}(\textbf{r}_1, \textbf{r}_2) + \psi_{ion}(\textbf{r}_1, \textbf{r}_2)$, where ionic and covalent parts are

\begin{align}
    \label{cov1}
&\psi_{cov}(\textbf{r}_1,\textbf{r}_2) =  \frac{2(t+V)}{\sqrt{2D(D-U+K)}}[
    w_1(\textbf{r}_1)w_2(\textbf{r}_2)\\ \nonumber &+w_1(\textbf{r}_2)w_2(\textbf{r}_1)] [\chi_{\uparrow}(1)\chi_{\downarrow}(2)-\chi_{\downarrow}(1)\chi_{\uparrow}(2)],
\end{align}
\noindent

\begin{align}
    \label{ion1}
&\psi_{ion}(\textbf{r}_1,\textbf{r}_2) = -\frac{1}{2}\sqrt{\frac{D-U+K}{\sqrt{2D}}}[ 
    w_1(\textbf{r}_1)w_1(\textbf{r}_2)\\ &+w_2(\textbf{r}_2)w_2(\textbf{r}_1)] 
    [\chi_{\uparrow}(1)\chi_{\downarrow}(2)-\chi_{\downarrow}(1)\chi_{\uparrow}(2)], \nonumber 
\end{align}
\noindent
with

\begin{align}
    D \equiv \sqrt{(U-K)^2+16(t+V)^2}, 
    \label{D}
\end{align}

\noindent
and

\begin{align}
    w_{i\sigma}(\textbf{r}) = \beta[\phi_{i\sigma}(\textbf{r})-\gamma
    \phi_{j\sigma}(\textbf{r})],
    \label{wannier}
\end{align}

\noindent
with $i=1$, $j=2$ or $i=2$, or $j=1$, in this case.
The two functions are molecular functions and come out naturally within our method, in which the two neighboring atomic functions $\phi_i(\textbf{r})$ are mixed, with $\beta$ and $\gamma$ as mixing parameters. These atomic functions 
can be in the form of Slater or Gaussian form (Slater or Gaussian type orbitals, STO or 
GTO). Furthermore, Eqs. \eqref{cov1} and \eqref{ion1} can be rewritten, with use of Eq. \eqref{wannier}, in 
the following way \cite{Hendzel1}

\begin{align}
    \label{cov2}
    &\psi_{cov}(\textbf{r}_1,\textbf{r}_2) =  \left(C\beta^2(1+\gamma^2) - 2\gamma I \beta^2\right)[\phi_1(\textbf{r}_1)
    \phi_2(\textbf{r}_2) \\ \nonumber &+\phi_2(\textbf{r}_1)\phi_1(\textbf{r}_2)] [\chi_{\uparrow}(1)\chi_{\downarrow}(2)-\chi_{\downarrow}(1)\chi_{\uparrow}(2)], 
\end{align}
\noindent
and
\begin{align}
 \label{ion2}
    &\psi_{ion}(\textbf{r}_1,\textbf{r}_2) = \left( I\beta^2(1-\gamma^2) - 2\gamma C \beta^2 \right)[\phi_1(\textbf{r}_1)
    \phi_1(\textbf{r}_2) \\ \nonumber &+\phi_2(\textbf{r}_1)\phi_2(\textbf{r}_2)] [\chi_{\uparrow}(1)\chi_{\downarrow}(2)-\chi_{\downarrow}(1)\chi_{\uparrow}(2)], 
\end{align}

\noindent
where $C$ and $I$ are coefficients from \eqref{cov1} and \eqref{ion1}, respectively. 

Parenthetically, for the sake of comparison one can write postulated VB two--particle wave function 
\begin{align}
    \label{cov3}
    &\psi^{VB}_{cov}(\textbf{r}_1,\textbf{r}_2) = \frac{1}{\sqrt{2(1+S^2)}}[\phi_1(\textbf{r}_1)
    \phi_2(\textbf{r}_2)+\phi_2(\textbf{r}_1)\phi_1(\textbf{r}_2)] \\ \nonumber
    & \times \frac{1}{\sqrt{2}}[\chi_{\uparrow}(1)\chi_{\downarrow}(2)-\chi_{\downarrow}(1)\chi_{\uparrow}(2)],
\end{align}
\noindent
and

\begin{align}
    \label{ion3}
    &\psi^{VB}_{ion}(\textbf{r}_1,\textbf{r}_2) = [\phi_1(\textbf{r}_1)
    \phi_1(\textbf{r}_2)+\phi_2(\textbf{r}_1)\phi_2(\textbf{r}_2)] \\ \nonumber
    &\times \frac{1}{\sqrt{2}}[\chi_{\uparrow}(1)\chi_{\downarrow}(2)-\chi_{\downarrow}(1)\chi_{\uparrow}(2)],
\end{align}

\noindent
where $S$ is the overlap between the neighboring atomic wave 
functions. However, the total wave function, consisting of 
sum of the \eqref{cov3} and \eqref{ion3} has not been 
obtained directly as a solution of the respective 
Schr\"odinger equation, whereas in our approach its form 
comes out explicitly from our exact solution and represents 
the exact treatment of the Heitler--London problem. 

Based on these functions we redefine the ionicity and covalency \cite{Hendzel2} and define atomicity \cite{Hendzel1}, the last is the complementary characteristic to the two former. 

A remark is in place at this point. As said above, he two--electron component wave functions \eqref{cov2} and \eqref{ion2} have formally the same form as their VB correspondents \eqref{cov3} and \eqref{ion3}, albeit with the two principal differences. First, the coefficients before the covalent and ionic parts, $\psi_{cov}$ and $\psi_{ion}$, are different as they contain \textbf{all} Coulomb--interaction terms between the particles composing the bond. Second, the orbital size ($\alpha^{-1}$) of the original atomic wave functions, composing those functions are adjusted in the resultant two--particle ground state. These two factors, in addition to the exact expression for the two--particle wave function, are the qualitative differences with the original Heitler--London theory.

In the next Section we discuss our results, after minimizing the ground state energy, $E_G[\psi_{G}(\alpha)] \equiv \braket{\psi_G(\alpha)|\mathcal{\hat{H}}|\psi_G(\alpha)}$/$\braket{\psi_G(\alpha)|\psi_G(\alpha)}$, with respect to $\alpha$ and evaluating explicitly the microscopic parameters for the optimal value of $\alpha=\alpha_0$.

\section{Results and discussion}

\begin{figure}
    \centering
    \includegraphics[width=0.5\textwidth]{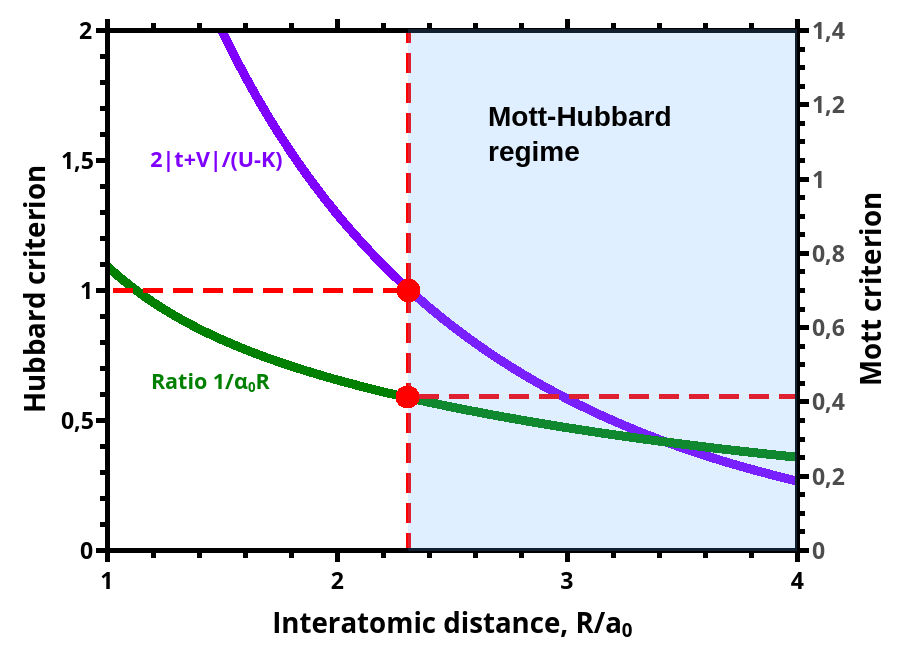}
    \caption{Mott (green, lower) and Hubbard (purple, upper) lines with the marked corresponding Mott and Hubbard criteria of localization. The shaded area to the right of $R=R_{Mott}$ represents the region with steadily increasing atomicity with increasing $R$. For details see main text.}
    \label{Fig1}
\end{figure}

\begin{figure}
    \centering
    \includegraphics[width=0.5\textwidth]{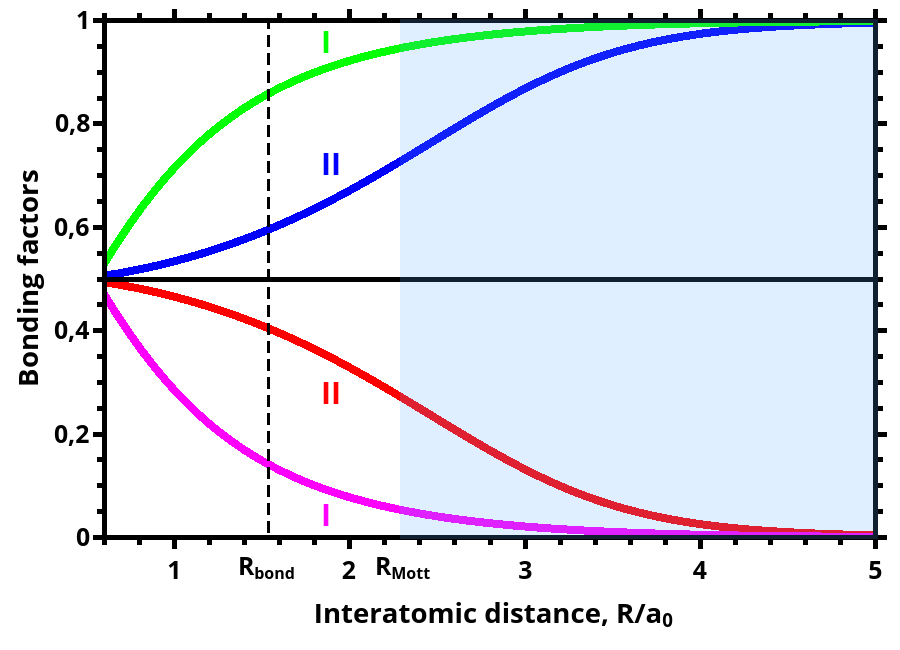}
    \caption{Comparison of starting binding factors (ionicity and standard covalency) vs. interatomic distance: Those microscopic parameters are determined, respectively: from Hubbard model – curves I; extended Hubbard model – curves II. The shaded area corresponds to the Mott--Hubbard (\textit{"Mottness"}) regime, where the interactions dominate over the kinetic energy particles (for details see Refs. \cite{Hendzel2, Hendzel1}).}
    \label{Fig2}
\end{figure}

We now proceed with the presentation of our results, followed up by their discussion. In Fig. \ref{Fig1} we illustrate the interatomic distance, $R$, dependence of the quantities with marked Hubbard and Mott criteria of localization (upper and lower red points, respectively). The Hubbard criterion (purple line) delineates the point where the kinetic to interaction ratio, $2|t+V|/(U-K)$, takes the value of unity. The Mottt criterion, in turn, describes the point where the atomic orbital size is of the same magnitude as the interatomic distance. The right--hand--side region (shaded) describes then the regime, where both the interaction dominates over the electron kinetic energy (according to the Hubbard criterion) and simultaneously, the atomic size in the correlated state is decisively smaller than the interatomic distance. Obviously, those criteria, crucial for the Mott-Hubbard localization in condensed matter, are only of qualitative nature in the case of molecules. They represent the finite--system situation and therefore, any sharp delocalization--localization transformation of molecular states into their atomic correspondents is ruled out.  Before discussing the details we show that the coefficients attached to the wave--function parts \eqref{cov1} and \eqref{ion1} represent the standard definition of covalency and ionicity, as is evident from the form of the corresponding component wave functions (second factors of the products in \eqref{cov2} and \eqref{ion2}, respectively). Their numerical values are displayed in Fig. \ref{Fig2}. For the sake of completeness, we have included in this Figure also results for the full solutions (curves labeled by I) and those corresponding to the Hubbard--model solutions (curves II). Parenthetically, the curves II describe the situation when we disregard all intersite Coulomb interaction and retain only the dominant term with intraatomic interaction $\sim U$. In either case, the covalency behaves unphysically with the interatomic distance $R \rightarrow \infty$ ($R>R_{\text{Mott}}$).

\begin{figure}
    \centering
    \includegraphics[width=0.5\textwidth]{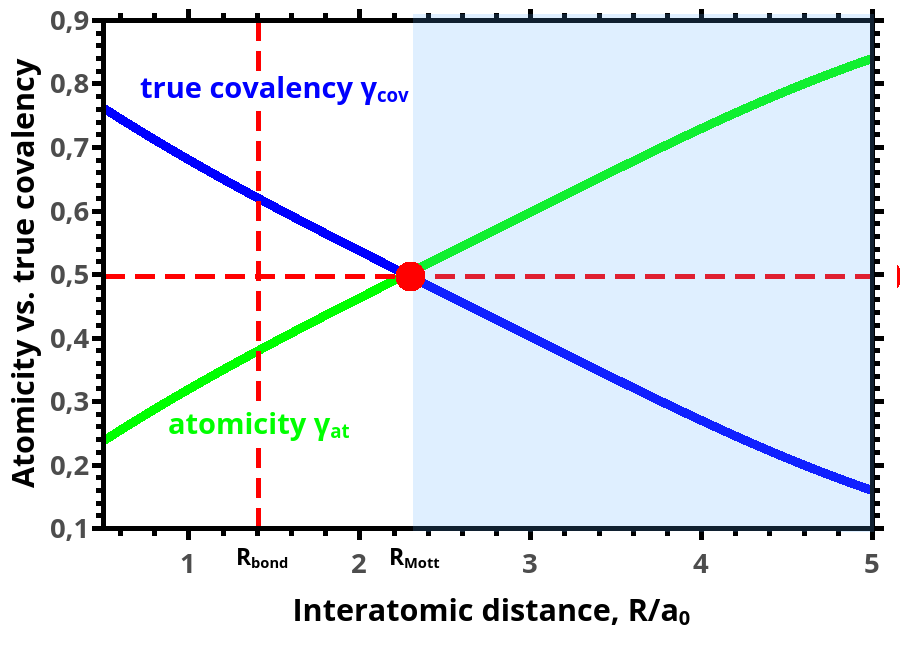}
    \caption{Participation of true covalency and atomicity in the resultant correlated state of electrons in \ch{H2}. Note that the two curves cross the marked point which corresponds accurately to the shown in Fig. \ref{Fig1}.}
    \label{Fig3}
\end{figure}

To restore physical meaning to the \textit{covalency} we make use of our earlier observation that in $R \rightarrow \infty$ limit the Heitler--London wave function reduces to the Slater determinant of the corresponding atomic states, with no ionicity, as it represents the probability amplitude of double occupancy on the same atom. We have proposed to exclude the \textit{atomicity} $\gamma_{\text{at}}$ from the covalency presented in Fig. \ref{Fig3} by extracting from the corresponding expression \eqref{cov2} the part taken for $\gamma = 0$ at given $R$ (not only in the atomic limit). As a result, we get the \textit{true covalency versus atomicity}, both as function of $R$, depicted in Fig. \ref{Fig3}. The ionicity remains without change, since it expresses the complementary factor of bonding --- the double occupancy. One should stress the fundamental difference between the covalency and ionicity factors, $\gamma_{\text{cov}}$ and $\gamma_{\text{ion}}$, shown in Fig. \ref{Fig2} and those exhibited in Fig. \ref{Fig3}. In the former case we have that                                                                       $\gamma_{\text{cov}}+\gamma_{\text{ion}} = 1$, whereas in the present situation $\gamma_{\text{cov}}+\gamma_{\text{ion}} + \gamma_{\text{at}} = 1$ (for details see \cite{Hendzel1}). It is remarkable, that the Mott--Hubbard criterion for localization meets the point where the defined atomicity $\gamma_{\text{at}}$ and redefined covalency $\gamma_{\text{cov}}$ are equal. Obviously, for larger $R$ values, the atomicity prevails, whereas the ionicity $\gamma_{\text{ion}}$ (not shown) decreases steadily to zero. 

\begin{figure}
    \centering
    \includegraphics[width=0.5\textwidth]{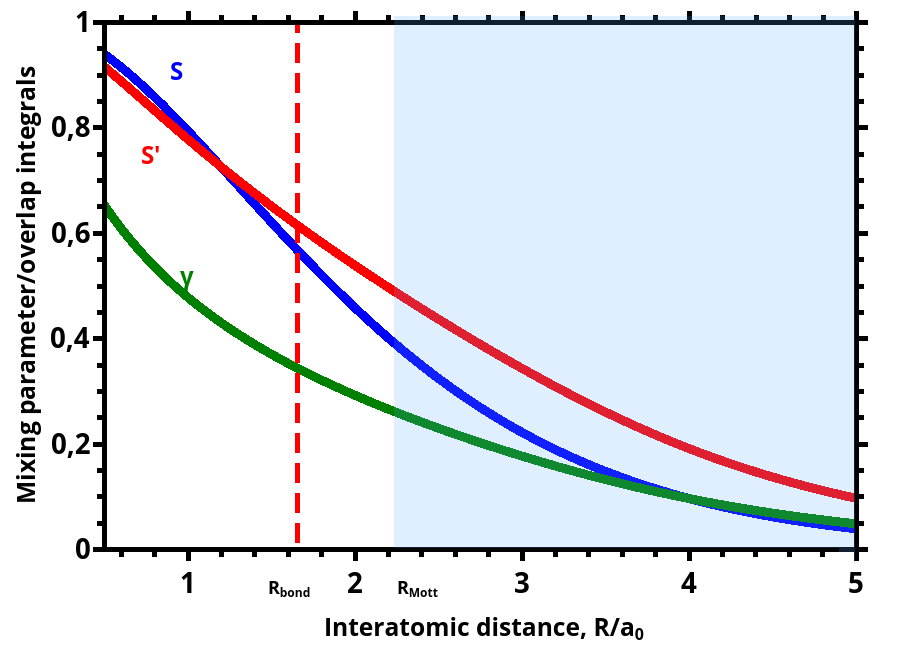}
    \caption{Single--particle mixing parameter $\gamma$, overlap integral without orbital size renormalization S, and orbital size calculated for renormalized orbital size S' versus interatomic distance R. They evolve continuously as a function of interatomic distance. The dashed vertical line marks, as all previous figures the equilibrium bond length, whereas the shaded area represent the \textit{"Mottness"} regime.}
    \label{Fig4}
\end{figure}

To illustrate our results by the way of showing that the onset of atomicity is a collective phenomenon, i.e., induced by electron--electron Coulomb interaction, we have plotted in Fig. \ref{Fig4} the single--particle characteristics $\gamma$ is admixture of neighboring and readjusted (in the correlated state) wave function, while $S$ and ${S}'$ are overlap integrals, for readjusted 
(${S}'$) and original $s$-state (${S}$) wave functions, respectively. All of those functions diminish gradually with the increasing $R$, without showing any sign of difference at either $R=R_{\text{Bond}}$ and $R=R_{\text{Mott}}$. In other words, the atomicity appears as a result of interelectronic interaction, induced by the correlations. 
To summarize, as well as to put our results in a broader prospective, we have listed selected properties of our calculations/computations in Tables \ref{Tab1} and \ref{Tab2}. There we have specified some standard quantities for the equilibrium state of \ch{H2} molecule (cf. Table \ref{Tab1}), as well as have singled out the bond characteristics (Table \ref{Tab2}). Additionally, we have supplemented these results with the true covalency, atomicity, and ionicity factors in Tables \ref{Tab3}/\ref{Tab4}. 
\begin{table}[]
\caption{Binding energy of \ch{H2} calculated with Restricted Hartree--Fock (RHF), Configurational Interaction (CI), and EDABI (with Hubbard Hamiltonian (HM--EDABI) and with extended Hamiltonian (EM--EDABI)) methods and percentage difference with the exact Ko\l{}os--Wolniewicz (K--W) result \cite{KW1968}. }
\begin{ruledtabular}
\begin{tabular}{ccc}
\hline
                                                                       & Binding energy (eV) & Difference with K--W ($\%$) \\ \hline
\begin{tabular}[c]{@{}c@{}}RHF\end{tabular}        & -3.5963     & 5.6     \\ \hline
\begin{tabular}[c]{@{}c@{}}Full CI\end{tabular}         &    -4.3824   & 0.6      \\ \hline
\begin{tabular}[c]{@{}c@{}}HM-EDABI\end{tabular}      &    -3.9783   &    3.1   \\ \hline
\begin{tabular}[c]{@{}c@{}}EM-EDABI\end{tabular} &    -4.0749       &    2.7     \\ \hline
\end{tabular}
\end{ruledtabular}
\label{Tab1}
\end{table}

\begin{table}[]
\caption{Bond length and correlation energy (calculated as $E_{HF}-E$ where $E_{HF}$ is Hartree-Fock energy and $E$ is energy in appropriate method) for \ch{H2} calculated with Restricted Hartree--Fock (RHF), Configurational Interaction (CI), and EDABI (with Hubbard Hamiltonian (HM--EDABI) and with extended Hamiltonian (EM--EDABI)) methods. }
\begin{ruledtabular}
\begin{tabular}{ccc}
\hline
                                                                       & Bond length ($a_0$) & Correlation energy (eV) \\ \hline
\begin{tabular}[c]{@{}c@{}}RHF\end{tabular}        & 1.450     & N/A     \\ \hline
\begin{tabular}[c]{@{}c@{}}Full CI\end{tabular}         &    1.501   & -0.5136     \\ \hline
\begin{tabular}[c]{@{}c@{}}HM-EDABI\end{tabular}      &    1.442   &    -0.0978     \\ \hline
\begin{tabular}[c]{@{}c@{}}EM-EDABI\end{tabular} &    1.430      &      -0.1706     \\ \hline
\end{tabular}
\end{ruledtabular}
\label{Tab2}
\end{table}

\begin{table}[]
\caption{Binding factors at $R = R_{Bond}$ calculated for many-particle wave function from extended second quantized Hamiltonian and Hubbard Hamiltonian as well as for single-particle wave function from Valence Bond Theory (with and without renormalizing orbital size).}
\begin{ruledtabular}
\begin{tabular}{ccc}
\hline
                                                                       & Covalency & Ionicity \\ \hline
\begin{tabular}[c]{@{}c@{}}Full Hubbard \\ Model\end{tabular}        & 0.59      & 0.41      \\ \hline
\begin{tabular}[c]{@{}c@{}}Hubbard \\ Model\end{tabular}         & 0.86      & 0.14      \\ \hline
\begin{tabular}[c]{@{}c@{}}Valence \\ Bond \\ Theory\end{tabular}      & 0.52      & 0.48      \\ \hline
\begin{tabular}[c]{@{}c@{}}Valence \\ Bond \\ Theory \\(renormalized)\end{tabular} & 0.63      & 0.37        \\ \hline
\begin{tabular}[c]{@{}c@{}}Space \\ Bonding \\ Descriptor \cite{Pendas}\end{tabular} & 0.57     & 0.43        \\ \hline
\end{tabular}
\end{ruledtabular}
\label{Tab3}
\end{table}

\begin{figure*}
    \centering
    \includegraphics[width=1\textwidth]{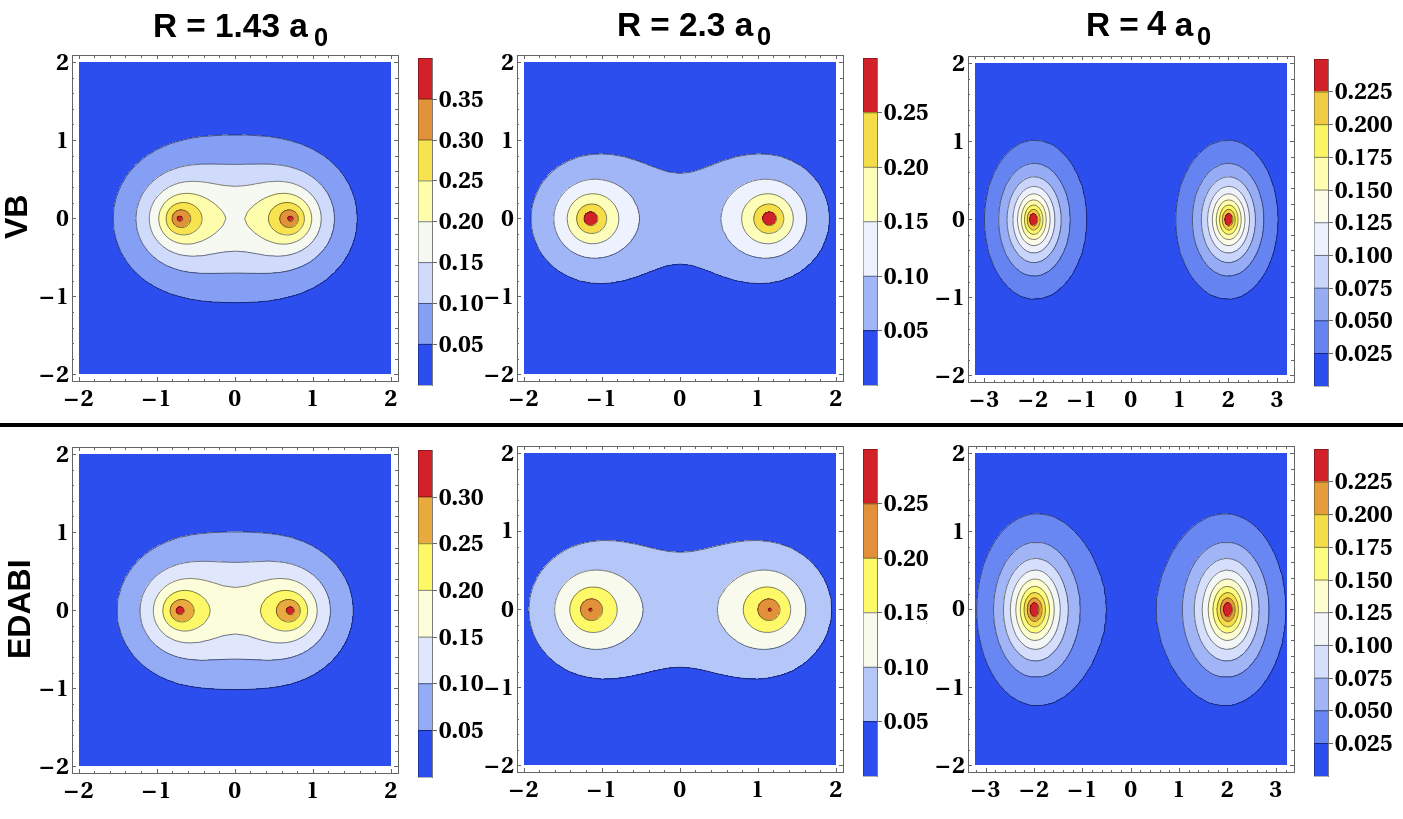}
    \caption{Probability density profiles according to \textbf{V}alence \textbf{B}ond (VB) approach --- upper panel, as well as those with taking the two--particle wave functions \eqref{cov1} and \eqref{ion1}. The interatomic distance $R$ is specified. }
    \label{Fig5}
\end{figure*}

\begin{table}[]
\caption{True binding characteristics for \ch{H2} at equilibrium point of $R$ with subtracted atomicity. Note that in Table \ref{Tab3} the atomicity is an integral part of the standard covalency.}
\begin{ruledtabular}
\begin{tabular}{ccc}
\hline
True covalency & Atomicity & Ionicity \\ \hline
\begin{tabular}[c]{@{}c@{}}  \end{tabular}    0.48    & 0.19    & 0.33    \\ \hline

\end{tabular}
\end{ruledtabular}
\label{Tab4}
\end{table}

%\begin{figure*}
%    \centering
%    \includegraphics[width=1\textwidth]{dens.png}
%    \caption{Comparison of electron probability calculated with VB and EDABI methods for three values of interatomic distance: $R = 1 \text{ }a_0$, bond length $R = 1.43 \text{ }a_0$, and Mott-Hubbard boundary $R = 2.3 \text{ }a_0$. }
%    \label{fig:dens}
%\end{figure*}

\section{Outlook}

The principal concept introduced in our approach 
\cite{Hendzel2} is the concept of \textit{atomicity} in 
nominally covalent bond of \ch{H2} (albeit, also with a 
nontrivial degree of ionicity). One should be aware of the 
fact that hydrogen molecule, in the hypothetical so far limit
$R \rightarrow \infty$, composed of separated atoms, is in an
incoherent quantum--mechanical state. Here we introduce such 
an incoherent admixture in the situation of still finite 
interatomic distance. This means that the entangled state of 
the two electrons in the correlated molecular state is then 
partially disentangled. One should still examine whether this
quantum--coherence limitation appears only for bound states, 
i.e., it is present also at finite distance when the 
particles interact, what is not the case with photons at any 
distance \cite{Aspect}. Until our last statement is proved, 
our proposal of atomicity in bound molecular states at finite
distance should be regarded as intuitive in nature, even 
though it helps to remove the principal inconsistency in 
evolution of the covalency as a function of interatomic 
distance. 

Finally, the overall behavior of the \ch{H2} system is illustrated in Fig. \ref{Fig5}, where the probability density profiles are shown for the three interatomic distances specified: At equilibrium bond distance ($R=1.43a_0$), at the Mott--Hubbard boundary ($R=2.3a_0$), and in the asymptotic regime of large distances ($R=4a_0$). The density profiles are artificially distorted from their almost spherical shapes by the choice of scale to expose the details of the density isolines in that case. One sees quasi--atomic character of the wave functions for $R>R_{Mott}$ in either approach, VB or EDABI. Nevertheless, the principal difference between the two is as follows. The limitation of the Heitler--London approach is caused by the selection of atomic $1s$ wave function to construct the two--particle state and the choice of the latter for the purely covalent state. In our case (EDABI) the single--particle states (molecular orbitals) with the adjusted size are taken; this effect is mainly due to the orbital size renormalization by their interaction. Additionally, the form of the two--particle wave function is more general as it contains also the ionic part. 

\section*{Acknowledgment}
This work was supported by Grants OPUS No.~UMO--2018/29/B/ST3/02646 and 
No.~UMO--2021/41/B/ST3/04070 from Narodowe Centrum Nauki. Critical remarks from Dr.~Maciek Fidrysiak are appreciated.

\bibliography{hendzel}
\end{document}